\documentclass[a4paper,12pt]{article}
\usepackage{epsfig}

\begin{document}

\begin{center}
{\bf VOLATILITY DYNAMICS OF WAVELET-FILTERED STOCK PRICES}\footnote{Supported by the RFBR grant 06-06-80357}
\end{center}

\bigskip

\begin{center}
{\bf I.M.~Dremin$^{(a)}$, A.V.~Leonidov$^{(a,b)\,}$\footnote{Corresponding author, e-mail leonidov@lpi.ru}}

\bigskip

\bigskip

(a) {\it Theoretical Physics Department, P.N.~Lebedev Physics Institute, \\ Moscow, Russia}

(b) {\it Institute of Theoretical and Experimental Physics, Moscow, Russia}

\end{center}

\bigskip

\bigskip

\begin{center}
{\bf Abstract}
\end{center}

Volatility dynamics of wavelet - filtered stock price time series is studied. Using the universal thresholding method of wavelet filtering and a
principle of minimal linear autocorrelation of noise component we find that the quantitative characteristics of volatility dynamics of denoised
series are noticeably different from those of the raw data and the noise.

\newpage

\section{Introduction}

Distinguishing valuable information from noise is one of the major problem one faces in developing a quantitative picture of price dynamics.
Generically one divides market participants into those making rational choices based on some adopted strategy and those making effectively random
decisions (noise traders). Understanding the impact of rational strategies requires therefore filtering of observed price series with respect to the
noise contribution. The main difficulty in separating meaningful signal from noise lies in the impossibility of giving a rigorous definition of what
noise is. Usual intuition visualizes noise as a high frequency uncorrelated dressing on top of a somewhat slower evolving signal containing useful
information. The problem of making this picture quantitative is twofold. First, noise can exist not just on some single scale, but on multiple
scales. In this case one should denoise the raw signal at several potentially relevant scales. Second, having noise with vanishing linear
autocorrelation does not imply that its higher order nonlinear autocorrelations are zero. A useful instrument for performing multiscale denoising is
a wavelet - based filtering, see e.g. \cite{DIN01,GSW01,R02}. As to the existence of high order nonlinear correlations in the noise component, this
will be a part of discussing volatility dynamics in the present paper. Let us also mention that recently wavelet - based denoising of financial time
series was discussed in \cite{BLT04,BLT06}, where a separation in the characteristics of active (turbulent) and inactive (laminar) periods in the
filtered signal was studied.

In the present paper we shall concentrate on studying volatity dynamics of wavelet-filtered financial series. Volatility (magnitude of price
increments) is one of the key characteristics of financial time series. In contrast with the classical diffusion (Brownian random walk) where
volatility is constant, financial time series are characterized by an intermittent pattern of high and low volatility periods. Such behavior points
to a presence of long-range memory of the magnitude of price increments (volatility), see e.g. \cite{BP03,Z03}. In particular, long memory property
of volatility dynamics shows itself in a very slow power-like decay of volatility autocorrelation\footnote{Detailed studies of volatility
correlations have revealed several characteristic time scales staying behind apparent long-range memory of volatity dynamics \cite{LB01,LZ03}.}.
Proper account for the impact of varying time horizons on volatility dynamics allows to construct a parsimonious model of stock price evolution
allowing to describe many important properties of observed stock price evolution \cite{LB01,BB05}.

In the present paper we focus on studying the autocorrelation properties of high-frequency volatility. We analyze the data for  5-minute returns of
the index MICEX10INDEX and five most liquid stocks EESR, RTKM, LKOH, SBER, SNGS traded at Moscow stock exchange MICEX during the years 2003-2005.

\section{Volatility dynamics of wavelet - filtered price series}

It is well - known that dynamical properties of stock price series are very close to those of the random walk. This refers in particular to a nearly
absent autocorrelation of stock price incremements. A more detailed analysis shows however, that zero autocorrelation does not mean that the
consequent price increments are independent. This dependence shows itself through more complex nonlinear correlators, e.g. those of lagged absolute
values of price increments.

For all the instruments under consideration we shall study a series of normalized 5 - minute logarithmic returns with zero mean and unit standard
deviation:
\begin{equation}
r^0_n = \log \left( \frac{p(t_n+\Delta T)}{p(t_n)} \right) \,\,\,\, => \,\,\,\,\, r_n = \frac{r^0_n-\langle r_n \rangle}{\sigma}
\end{equation}

Filtering of the original series $ \{ r \}$ corresponds to its separation into filtered and noise components:
\begin{equation}
 r \, = \, r_F + r_N
\end{equation}

As mentioned in the Introduction, denosing financial time series is a crucial step in unraveling the probabilistic dependence patterns hidden in it.
The main problem is that a rigorous definition of noise based on some fundamental principles of theoretical finance does not exist. Therefore when
considering a problem of filtering noise from a series an operational definition of noise should be formulated. In financial applications the most
important characteristic of noise is, ideally, an absence of any type of predictability corresponding to complete probabilistic independence of the
terms in the series under consideration. In practice it is practically impossible to have a denoising procedure ensuring that all nonlinear
autocorrelation functions are zero. We shall therefore restrict our consideration to the usual linear autocorrelation function with lag $1$ and
require that the lagged autocorrelation of noise component of price increments $\rho_N (1)$ vanishes:
\begin{equation}\label{acfn}
 \rho_N(1) \equiv \langle r_N(t) \, r_N(t+\Delta T) \rangle = 0
\end{equation}
The property (\ref{acfn}) ensures that the noise component does not contain linear predictability, i.e. that a conditional mean $\langle r_N(t+\Delta
T) \rangle_{r_N(t)}= \rho_N \cdot r(t)$ vanishes for any $r_N(t)$ so that noise does not contain trivial profitable opportunities.

The filtering procedure we employ is based on the universal threshold wavelet filtering \cite{DJ93a,DJ93b}. It is realized in three stages:
\begin{itemize}
\item{
        Wavelet transformation of the original time series of price returns $r(t)$ with discrete wavelet transformation (DWT):
        \begin{equation}
        r(t) \Longrightarrow w_{j,n}=2^{-j/2} \int \, dt \, r(t) \, \psi (2^{-j} t - n)
        \end{equation}
        where $\psi(t)$ is a mother wavelet function which we will choose as that of the Daubechy 2 wavelet.
     }
\item{
        Separating the set of wavelet coefficients $w_{j,n}$ into the noise  $w_{j,n}^N$  and signal $w_{j,n}^F$ contributions through
        universal thresholding procedure \cite{DJ93a,DJ93b} where the signal components satisfy, at each resolution level $j$, the
        requirement
        \begin{equation}
        |\,w_{j,n}^F| > \sqrt{2 \log n} \, \sigma_{j,n} \, ,
        \end{equation}
        $n$ is a number of wavelet coefficients at level $j$ and $\sigma_{j,n}$ is their standard deviation.
     }
\item{
        Inverse wavelet transformation of the set $\{w_{j,n}^F \}$ producing the filtered time series of returns $r^F(t)$ and, by default, a
        noise component $r^N(t) \equiv r(t)-r^F(t)$.
     }
\end{itemize}
As described above, to completely specify the optimal filtering procedure we need to give an operational definition of noise. Below we shall define
optimal filtering as that leading to the smallest 1-lag autocorrelation coefficient of noise component within the adopted filtering procedure. A
natural parameter  within the universal thresholding scheme one can tune is a number of levels involved in thresholding process. The considered time
series consist of $16384=2^{14}$ 5-minute intervals, so one has 14 levels of resolution. A customary choice is applying thresholding to all levels
starting from the third. We have considered a starting level as a tunable parameter allowing to achieve optimal (i.e. leading to smallest one-lag
autocorrelation coefficient of the noise component) filtering. An analysis showed that for the index MICEX10INDEX it was optimeal to start with the
7-th level, whereas for all five stocks the optimal starting level is the 3-d. An illustration of the effect of filtering is given in Fig.~1 in which
the raw cumulative normalized returns and their wavelet-filtered counterparts for the index MIXEX10INDEX and the stock LKOH are plotted. In Fig.~1 we
clearly see a difference that a choice of the starting level makes: filtering of the index series starts from a much finer resolution than that of a
stock.

Let us now compare some properties of the raw and filtered time series $r(t)$ and $r^F(t)$. Two most interesting questions are:
\begin{itemize}
\item{Does filtering induce linear autocorrelation in the filtered time series?}
\item{Is the filtered time series more/less non-gaussian than the original one?}
\item{What are the correlation patterns characterizing the filtered series as compared to those of noise and original noisy series?}
\end{itemize}
The answer is the first question is given by the lagged autocorrelation function $\rho(l)$. To answer the second one one has to introduce a
quantitative measure of the distance between the distributions of returns ${\cal P} (r)$ and  ${\cal P} (r^F)$ and their gaussian distributions
having the same standard deviation. A convenient quantitative measure of this distance is the (normalized) hypercumulant \cite{BP03}:
\begin{equation}
 \kappa = \sqrt{\frac{\pi}{2}} \, \frac{\langle |\,r| \rangle}{\sigma_r}
\end{equation}
For the gaussian distribution $\kappa = 1$, whereas $\kappa <1 $ means that the corresponding distribution is leptokurtic (i.e. having positive
anomalous kurtosis and heavy tails). The corresponding characteristics for all instruments under consideration are shown in Table1:

\begin{center}
{\bf Table 1}

\bigskip

\begin{tabular}{|c|c|c|c|c|}
  \hline
   {\bf Instrument} & {\bf $\bf \rho (1)$} & {\bf $\rho_F (1)$} & {$\bf \kappa$} & {\bf $\kappa_F$} \\ \hline
  MICEX10IND & -0.0036 & -0.0512 & 0.7872 & 0.3178 \\ \hline
  EESR & 0.0013 & 0.0152 & 0.8038 & 0.203 \\ \hline
  RTKM & -0.0431 & -0.0648 & 0.8533 & 0.1773 \\ \hline
  LKOH & -0.0128 & -0.0246 & 0.8641 & 0.2089 \\ \hline
  SNGS & -0.0602 & -0.2872 & 0.8856 & 0.19 \\ \hline
  SBER & -0.0755 & -0.2618 & 0.835 & 0.1747 \\ \hline
\end{tabular}
\end{center}

\medskip

We see that the autocorrelations of the original series are somewhat amplified in the filtered ones but remain at the insignificant levels (except
for SNGS and SBER). A much more pronounced effect is seen in the drastic difference between the values of normalized hypercumulant $\kappa$ in hte
raw abd filtered series. We see that the filtered series are much more non-gaussian than the raw ones.

To address the third question let us now turn to the analysis of the volatility dynamics of raw, wavelet-transformed and noise time series. We will
use a simple definition of volatility as an absolute value of price returns. A fundamental characteristics of volatility dynamics by is a lagged
autocorrelation function
\begin{equation}
g(k) \, = \, \frac{\langle (|\,r_n| -\langle |\,r| \rangle ) (|\,r_{n+k}| - \langle |\,r| \rangle ) \rangle}{\sigma^2 (|\,r|)}
\end{equation}
One of the most important properties of financial time series is a slow powerlike decay of $g(k)$ with $k$ showing that a stochastic process
governing temporal evolution of volatility is a long-range memory one, see e.g. \cite{BP03}.

The volatility autocorrelation functions of the raw, filtered and noise series for the index MICEX10INDEX and the corresponding averaged volatility
autocorrelations of five stocks under consideration are shown in Fig.~2. All of them indeed demonstrate a clear powerlike behavor
\begin{equation}
g(k) \, = \, g_0 \,  k^{-\alpha}
\end{equation}
The slopes $\{ \alpha \}$ of the powerlike fits to the curves shown in Fig.~2 are given in Table 2:

\begin{center}
{\bf Table 2}

\bigskip

\begin{tabular}{|c|c|c|c|}
  \hline
  {\bf Instrument} & Raw & Filtered & Noise \\ \hline
  MICEX10INDEX & -0.19 & -0.52 & -0.13 \\ \hline
  Stocks & -0.19 & -0.68 & -0.18 \\ \hline
\end{tabular}
\end{center}

 A striking feature of the plots in Fig. 2 and, correspondingly, the values in Table 2 is that the
while autocorrelation properties of the raw and noise series are practically the same, the filtered series has approximately a two times larger
amplitude $g_0$ and a steeper decay with $\alpha=0.67$ for stocks and $\alpha=0.52$ for the index compared to the slope of $\alpha = 0.19$ for both
of the original time series. This shows that wavelet denoising reveals volatility dynamics very different from that seen in the noisy orignal
series\footnote{Let us mention that the sensitivity of calibrating volatility dynamics to noise effects is very large, see e.g. the comment in
\cite{BB05}.}. This property can be very important in terms of examining real degree of predictability in volatility dynamics which is of great
interest in many practical applications. This fact constitutes a main result of the present paper

\section{Conclusions and outlook}

Let us formulate once again the main conclusions of the present study. We used a procedure of filtering the original price return series by the
universal thresholding of the coefficients of its discrete wavelet transform. We found that
\begin{itemize}
\item{The filtered series is much more non-gaussian as the original one.}
\item{The filtered series is characterized by a drastically different volatility autocorrelation function with larger amplitude and slope than
in the original series. }
\item{The volatility autocorrelation function of noise is very close to that of the initial series.}
\end{itemize}

In the present paper we considered only one particular characteristics of the wavelet - filtered - its volatility autocorrelation functions. Recent
studies have revealed a number of striking dependence patterns characterizing high frequency dynamics of stock prices
\cite{LTZ05,LTZZ06a,LTZZ06b,LTZZ06c}. It will be very interesting to see how these dependence patterns change when one considers the denoised price
series. Work on these issues is currently in progress.

\begin{center}
{\bf Acknowledgements }
\end{center}

The work was supported by RFBR grant 06-06-80357.

\begin{figure}[h]
\begin{center}
\epsfig{file=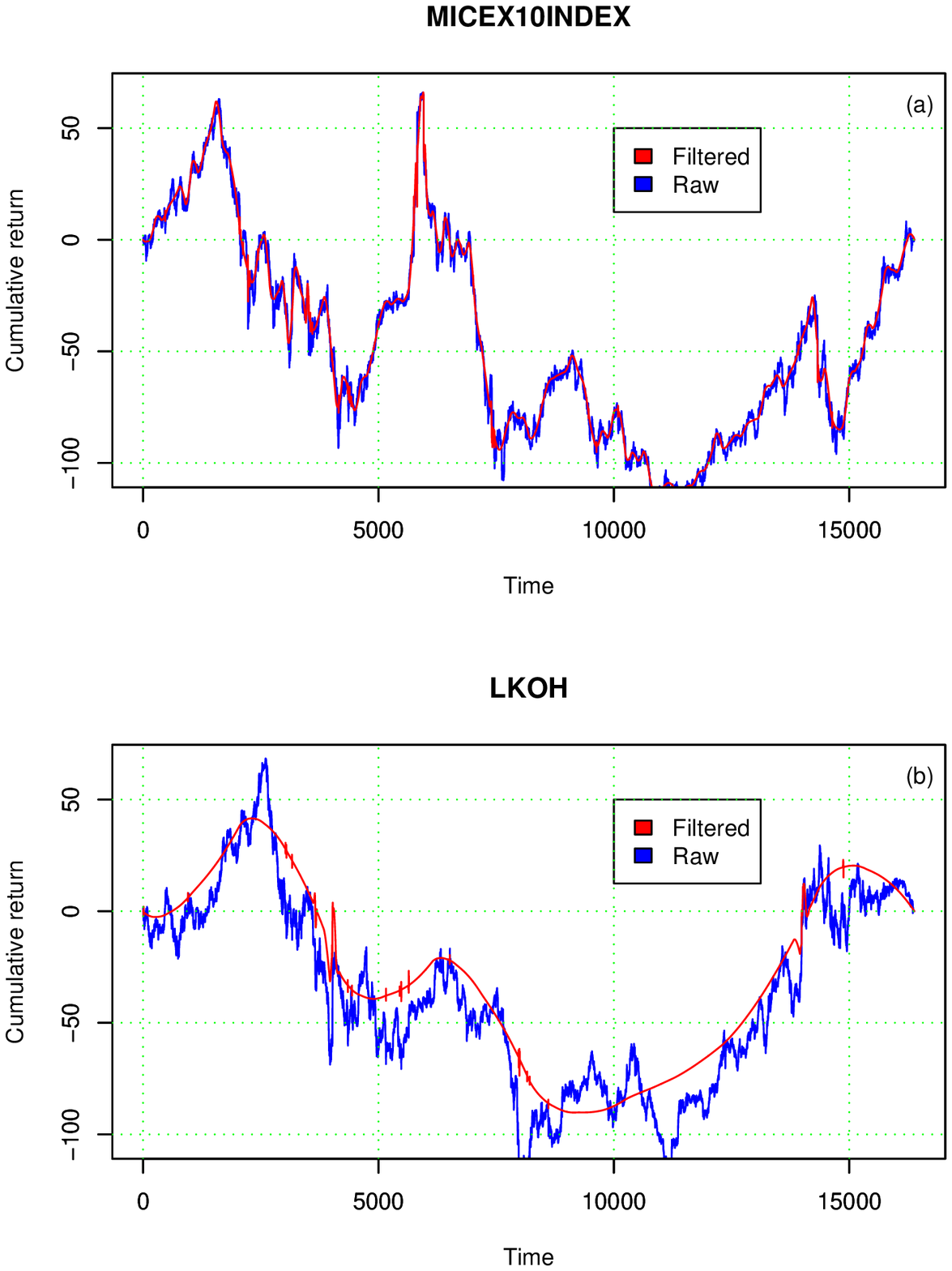,height=18cm,width=16cm}
\end{center}
\caption{Raw versus wavelet-filtered price dynamics for (a) MICEX10INDEX; (b) LKOH} \label{volacor}
\end{figure}

\begin{figure}[h]
\begin{center}
\epsfig{file=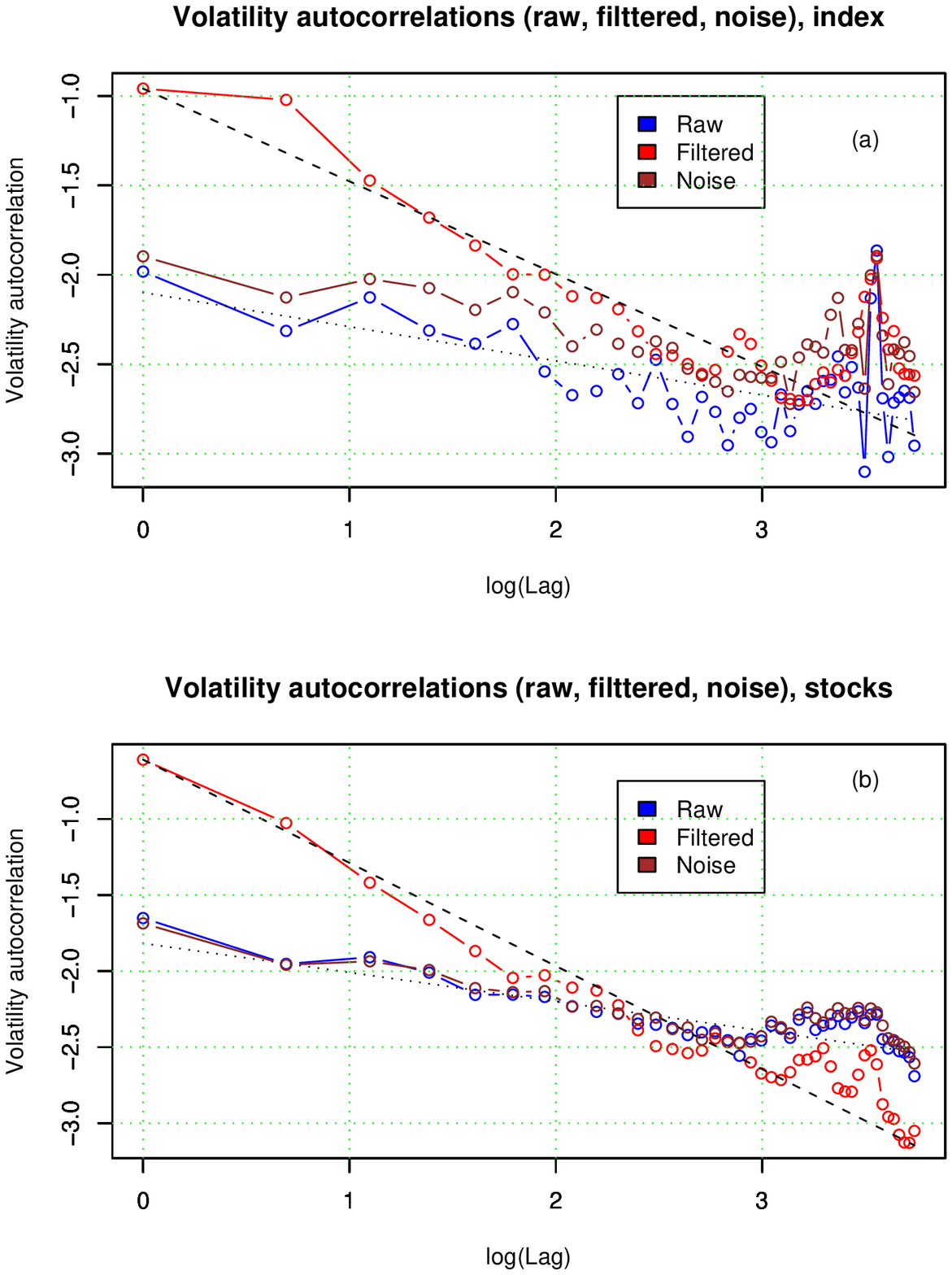,height=18cm,width=16cm}
\end{center}
\caption{(a) Volatility atocorrelations for filtered, original and noise series for MICEX10INDEX.
          (b) Average volatility atocorrelations for filtered, original and noise series for stocks.
          Dashed lines: powerlike fits for filtered autocorrelation; dotted lines: powerlike fits for raw autocorrelations.  } \label{volacor}
\end{figure}


\begin{thebibliography}{99}

\bibitem{DIN01}
I.M.~Dremin, O.V.~Ivanov, V.A.~Nechitailo, "Wavelets and their use", {\it Physics-Uspekhi}\, {\bf 44} (2001), 447-552

\bibitem{GSW01}
R.~Gencay, F.~Selcuk, B.~Whitcher, "An Introduction to Wavelets and other Filtering Methods in Finance and Economics", Springer 2001

\bibitem{R02}
J.B.~Ramsey, "Wavelets in Economics and Finance: Past and Future", C.V. Starr Center for Applied Economics report RR 2002-02

\bibitem{BP03}
J.-P.~Bouchaud, M.~Potters, "Theory of Financial Risk and Derivative Pricing", Oxford Univeristy Press, 2003

\bibitem{Z03}
G.~Zumbach, "Volatility processes and volatility forecast woth long memory", {\it Quantitative Finance}\, {\bf 4} (2004), 70-86

\bibitem{LB01}
B.~LeBaron, "Stochastic Volatility as a Simple Generator of Financial Power-laws and Long Memory", {\it Quantitative Finance}\, {\bf 1} (2001), 631

\bibitem{LZ03}
P.E.~Lynch, G.O.~Zumbach, "Market heterogeneities and the causal structure of volatility", {Quantitative Finance}\, {\bf 3} (2003), 320-331

\bibitem{BB05}
L.~Borland, J.-P.~Bouchaud, "On a multi-timescale statistical feedback model for volatility correlations", [ArXiv:physics/0507073]

\bibitem{BLT04}
M. Bartolozzi, D.B. Leinweber and A.W. Thomas, "Self-Organized Criticality and Stock Market Dynamics: an Empirical Study", [ArXiv:cond-mat/0405257]

\bibitem{BLT06}
M. Bartolozzi, D.B. Leinweber and A.W. Thomas, "Scale-free avalanche dynamics in the stock market", [arXiv:physics/0601171]

\bibitem{DJ93a}
D. Donoho, I. Johnstone, "Ideal spatial adaptation by wavelet shrinkage", {\it Biometrica} (1993)

\bibitem{DJ93b}
D. Donoho, I. Johnstone, "Adapting to unknown smoothness via wavelet shrinkage", {\it Journ.\ Am.\ Stat.\ Ass.} (1993)

\bibitem{LTZ05}
A.~Leonidov, V.~Trainin, A.~Zaitsev, "On collective non-gaussian dependence patterns in high frequency financial data", ArXiv:physics/0506072

\bibitem{LTZZ06a}
A.~Leonidov, V.~Trainin, A.~Zaitsev, S.~Zaitsev, "Market Mill Dependence Pattern in the Stock Market: Asymmetry Structure, Nonlinear Correlations and
Predictability", arXiv:physics/0601098.

\bibitem{LTZZ06b}
A.~Leonidov, V.~Trainin, A.~Zaitsev, S.~Zaitsev, "Market Mill Dependence Pattern in the Stock Market: Distribution Geometry, Moments and
Gaussization", arXiv:physics/0603103.

\bibitem{LTZZ06c}
A.~Leonidov, V.~Trainin, A.~Zaitsev, S.~Zaitsev, "Market Mill Dependence Pattern in the Stock Market: Distribution Geometry. Individual Portraits",
arXiv:physics/0605138.

\end{thebibliography}
\end{document}